\documentclass[]{jetp}

\begin{document}
\English

\title{Drag force and superfluidity in the supersolid stripe phase of a spin-orbit-coupled Bose-Einstein condensate}

\author{G.~I.}{Martone}
\email{giovanni.martone@u-psud.fr}
\affiliation{LPTMS, CNRS, Univ. Paris-Sud, Universit\'{e} Paris-Saclay, Orsay 91405, France}

\author{G.~V.}{Shlyapnikov}
\affiliation{Russian Quantum Center, Skolkovo, Moscow 143025, Russia}
\affiliation{SPEC, CEA, CNRS, Universit\'{e} Paris-Saclay, CEA Saclay, Gif sur Yvette 91191, France}
\affiliation{LPTMS, CNRS, Univ. Paris-Sud, Universit\'{e} Paris-Saclay, Orsay 91405, France}
\affiliation{Van der Waals-Zeeman Institute, Institute of Physics, University of Amsterdam, Science Park 904, 1098 XH Amsterdam, The Netherlands}
\affiliation{Wuhan Institute of Physics and Mathematics, Chinese Academy of Sciences, 430071 Wuhan, China}
\affiliation{Russian Quantum Center, National University of Science and Technology MISIS, Moscow 119049, Russia}

\abstract{The phase diagram of a spin-orbit-coupled two-component Bose gas includes a supersolid stripe phase, which is featuring density modulations along the direction
of the spin-orbit coupling. This phase has been recently found experimentally [J.~Li \textit{et al.}, Nature (London) \textbf{543}, 91 (2017)]. In the present work we characterize
the superfluid behavior of the stripe phase by calculating the drag force acting on a moving impurity. Because of the gapless band structure of the excitation spectrum,
the Landau critical velocity vanishes if the motion is not strictly parallel to the stripes, and energy dissipation takes place at any speed. Moreover, due to the spin-orbit coupling,
the drag force can develop a component perpendicular to the velocity of the impurity. Finally, by estimating the time over which the energy dissipation occurs, we find that
for slow impurities the effects of friction are negligible on a time scale up to several seconds, which is comparable with the duration of a typical experiment.}

\maketitle

\section{Introduction}
\label{sec:introduction}
Supersolidity is an intriguing phenomenon characterized by a simultaneous presence of superfluidity and crystalline order (see~\cite{Boninsegni2012review} for a review).
The existence of supersolid states was originally proposed by Gross~\cite{Gross1957,Gross1958}. Subsequently, these states were discussed in a general context and in
the context of helium in pioneering works by Andreev and Lifshitz~\cite{Andreev1969}, Leggett~\cite{Leggett1970}, Chester~\cite{Chester1970}, Kirzhnits and Nepomnyashchii~\cite{Kirzhnits1971},
and Pitaevskii~\cite{Pitaevskii1984}. The possibility of achieving the supersolid state in solid helium has been the subject of a long-lasting debate, but no incontrovertible
experimental evidence in such a system has been brought up to now~\cite{Balibar2010}. In the dilute limit for bosons in free space, supersolidity appears as a Bose-Einstein condensate
(BEC) with the wavefunction that has the form of a crystal lattice on top of a uniform background. The supersolid behavior has been theoretically predicted in ultracold atomic gases,
notably in configurations with soft-core two-body potentials~\cite{Pomeau1994,Henkel2010,Cinti2010,Saccani2011} and in two-dimensional dipolar BECs~\cite{Buchler2007,Astrakharchik2007,Kurbakov2010,Lu2015}.

In the last few years, supersolidity has been investigated in BECs with spin-orbit (SO) coupling. The latter arises, in particular, by coupling two spin (more precisely, 
pseudospin) degrees of freedom of a BEC through a pair of  Raman beams, and the single-particle dispersion may then feature multiple minima at finite momenta. Supersolidity
originates from the interplay between two-body interaction and modified single-particle dispersion. Stripe phases with supersolid properties have been studied in several kinds
of SO-coupled configurations~\cite{Wang2010,Ho2011,Wu2011,Li2012PRL} (see also reviews~\cite{Galitski2013review,Zhou2013review,Zhai2015review,Li2015review,Zhang2016review}
and references therein for a wider overview of SO-coupled quantum gases). Their observation, however, has remained an open problem for a long time after the first
experimental realization of a BEC with SO coupling by the NIST group~\cite{Lin2011}. The implementation of a new setup based on optical superlattices at MIT~\cite{Li2016}
has allowed to overcome the main limitations of the previous works, i.e. the smallness of the contrast of the density modulations and the high sensitivity to magnetic
fields~\cite{Martone2014,Martone2015}. This eventually has led to the detection of the stripe phase by Bragg spectroscopy~\cite{Li2017}. Together with the experiment
of the ETH group~\cite{Leonard2017} that employed a BEC coupled to two optical cavities, the MIT experiment~\cite{Li2017} represents the first observation of supersolidity
in ultracold gases.

The excitation spectrum of the stripe phase, which was theoretically investigated in Ref.~\cite{Li2013}, is of particular interest. It features a band structure with two gapless
branches, corresponding to the Goldstone modes associated with the spontaneously broken gauge and translational invariance. The frequency of these two lowest-lying modes
vanishes at the edge of the first Brillouin zone. Such a band structure implies a vanishing Landau critical velocity. This appears to be consistent with the findings of Ref.~\cite{Pomeau1994},
where the authors showed that the model of supersolid that they considered cannot support dissipationless flow around an obstacle at any velocity. It is thus natural to wonder
how to characterize the superfluid behavior in such situations. In this paper we discuss this problem for the stripe phase of a SO-coupled BEC by deriving the drag force
in the defect propagation. We follow the procedure, which was originally developed by Astrakharchik and Pitaevskii~\cite{Astrakharchik2004}. It is based on the evaluation of the
linear response of the BEC to a $\delta$-function perturbation, corresponding to a localized heavy impurity moving with a constant velocity. From this we deduce the time scale
over which the energy dissipation occurs. As we will see, for parameters similar to those of Ref.~\cite{Li2017} this time scale is fairly large, ranging
from a few tenths of a second to several seconds. We thus conclude that the motion of a slow body through the gas can be considered as dissipationless for the whole
duration of the experiment.

The paper is organized as follows. In Sec.~\ref{sec:model} we describe the SO-coupled model that we are considering. Section~\ref{sec:stripe_phase} is devoted to a brief
review of the static and dynamic properties of the stripe phase. The calculation of the drag force and of the time scale for the energy dissipation is presented
in Section~\ref{sec:drag_force}. We conclude in Sec.~\ref{sec:conclusions}. In the appendix we give additional details on the computation of the drag force.

\section{The model}
\label{sec:model}
We consider a three-dimensional (3D) two-component BEC (spin-$1/2$ bosons) featuring an equal-weighted superposition of Rashba~\cite{Bychkov1984} and
Dresselhaus~\cite{Dresselhaus1955} SO couplings. The single-particle Hamiltonian reads
\begin{equation}
h_{\mathrm{SO}} = \frac{\left(p_x - \hbar k_R \sigma_z\right)^2}{2m} + \frac{p_\perp^2}{2m} + \frac{\hbar\Omega_R}{2} \, \sigma_x + \frac{\hbar\delta_R}{2} \, \sigma_z \, .
\label{eq:sp_hamiltonian}
\end{equation}
This system has been realized experimentally for the first time in Ref.~\cite{Lin2011}, and subsequently it has represented the framework in which the supersolid stripe phase
has been observed~\cite{Li2017}. Hamiltonian~(\ref{eq:sp_hamiltonian}) is produced by coupling two spin atomic states (actually, these are pseudospin states, but in the
following we will keep the term spin for brevity) through a pair of Raman beams. The strength of the SO coupling is fixed by the momentum transfer due to the lasers.
It is equal to $\hbar k_R \hat{\mathbf{e}}_x$, where $\hat{\mathbf{e}}_x$ is the unit vector along the direction $x$. The Raman coupling $\hbar \Omega_R$ is instead related
to the intensity of the light field. The quantities $m$ and $\sigma_{x,y,z}$ denote the atom mass and the $2 \times 2$ Pauli matrices, while $p_\perp^2 = p_y^2 + p_z^2$.
The effective Zeeman shift $\hbar \delta_R$ quantifies the detuning (from the Raman resonance) of the transition between the two spin states, and we take
$\delta_R = 0$.

Hamiltonian~(\ref{eq:sp_hamiltonian}) is static and translationally invariant. Hence, one has a complete set of eigenstates in the form of plane waves with momentum
$\mathbf{p}$. The single-particle energy spectrum includes two branches:
\begin{equation}
\varepsilon_{\pm}(\mathbf{p}) = \frac{\mathbf{p}^2}{2m} + E_R \pm \sqrt{\left( \frac{\hbar k_R p_x}{m} \right)^2 + \left(\frac{\hbar\Omega_R}{2}\right)^2} \, ,
\label{eq:sp_spectrum}
\end{equation}
with $E_R = (\hbar k_R)^2/2m$ being the Raman recoil energy. The single-particle ground state is identified by the minima of the lower branch. It turns out that, in the
$\hbar\Omega_R < 4 E_R$ regime, $\varepsilon_-(\mathbf{p})$ has two degenerate minima located at finite momenta $\mathbf{p} = \pm \hbar \mathbf{k}_1^0
= \pm \hbar k_1^0 \hat{\mathbf{e}}_x$, where
\begin{equation}
k_1^0 = k_R \sqrt{1 - \left(\frac{\hbar\Omega_R}{4 E_R}\right)^2} \, .
\label{eq:k1_sp}
\end{equation}
If instead $\hbar\Omega_R \geq 4 E_R$, then the lower branch has a single minimum at $\mathbf{p} = 0$.

Let us now assume that the condensate is in a volume $V$ and has $N$ particles, which interact with each other via a two-body contact potential. In second quantization
the system is described by introducing a two-component field operator $\hat{\Psi}(\mathbf{r}) = ( \hat{\Psi}_{\uparrow}(\mathbf{r}) \,\, \hat{\Psi}_{\downarrow}(\mathbf{r})  )^T$
obeying the usual bosonic commutation rules ($T$ is the transposition symbol). The many-body Hamiltonian can be written as
\begin{equation}
\hat{H} = \int_V d^3 r \left\{ \hat{\Psi}^\dagger(\mathbf{r}) h_{\mathrm{SO}} \hat{\Psi}(\mathbf{r})
+ \frac{g_{dd}}{2} \hat{n}^2(\mathbf{r}) + \frac{g_{ss}}{2} \hat{s}_z^2(\mathbf{r}) \right\} \, ,
\label{eq:H}
\end{equation}
where $\hat{n}(\mathbf{r}) = \hat{\Psi}^\dagger(\mathbf{r}) \hat{\Psi}(\mathbf{r})$ is the total density, and $\hat{s}_z(\mathbf{r})
= \hat{\Psi}^\dagger(\mathbf{r}) \sigma_z \hat{\Psi}(\mathbf{r})$ is the spin density along the direction $z$ (normal ordering of the quantum fields $\hat{\Psi}$ and $\hat{\Psi}^\dagger$
is implied in Eq.~(\ref{eq:H})). The two interaction strengths are given by $g_{dd} = (g+g_{\uparrow\downarrow}) / 2$ and $g_{ss} = (g-g_{\uparrow\downarrow}) / 2$,
where $g$ and $g_{\uparrow\downarrow}$ are the intraspecies and interspecies coupling constants, respectively (we assume $g_{\uparrow\uparrow} = g_{\downarrow\downarrow} = g$). The coupling constants are related to the corresponding $s$-wave scattering lengths
$a_{\sigma\sigma'}$ as $g_{\sigma\sigma'} = 4\pi \hbar^2 a_{\sigma\sigma'} / m$ ($\sigma,\sigma' = \uparrow,\downarrow$). In the Heisenberg picture
the time evolution of the quantum field $\hat{\Psi}(\mathbf{r},t)$ is governed by the Heisenberg equation
\begin{equation}
i\hbar\partial_t \hat{\Psi}(\mathbf{r},t) = [\hat{\Psi}(\mathbf{r},t),\hat{H}]
= \left[ h_{\mathrm{SO}} + g_{dd} (\hat{\Psi}^\dagger(\mathbf{r},t) \hat{\Psi}(\mathbf{r},t))
+ g_{ss} (\hat{\Psi}^\dagger(\mathbf{r},t) \sigma_z \hat{\Psi}(\mathbf{r},t)) \sigma_z\right] \hat{\Psi}(\mathbf{r},t) \, .
\label{eq:heisemb_eq}
\end{equation}

The properties of our interacting SO-coupled BEC can be investigated by using the Gross-Pitaevskii (GP) mean-field approach, which allows one to determine
equilibrium configurations. Small oscillations around these configurations can be studied using the Bogoliubov theory. For this purpose, we decompose the field operator as
\begin{equation}
\hat{\Psi}(\mathbf{r},t) = e^{- i \mu t / \hbar} \left[ \Psi_0(\mathbf{r}) + \delta \hat{\Psi}(\mathbf{r},t) \right] \, .
\label{eq:field_decomp}
\end{equation}
Here $\Psi_0(\mathbf{r})$ is a classical field (condensate wavefunction), and $\delta \hat{\Psi}(\mathbf{r},t)$ characterizes small fluctuations on top of the equilibrium state. The quantity $\mu$ is the chemical
potential, which is fixed by the normalization condition $\int_V d^3 r \, \Psi_0^\dagger(\mathbf{r}) \Psi_0(\mathbf{r}) = N$. Let us now
insert Ansatz~(\ref{eq:field_decomp}) into Eq.~(\ref{eq:heisemb_eq}) and retain only terms up to the first order in $\delta \hat{\Psi}(\mathbf{r},t)$. One then finds that $\Psi_0$
obeys the stationary GP equation~\cite{Pitaevskii_Stringari_book,Pethick_Smith_book}:
\begin{equation}
\left[h_{\mathrm{SO}} + g_{dd} (\Psi_0^\dagger(\mathbf{r}) \Psi_0(\mathbf{r})) + g_{ss} (\Psi_0^\dagger(\mathbf{r}) \sigma_z \Psi_0(\mathbf{r})) \sigma_z\right] \Psi_0(\mathbf{r})
= \mu \Psi_0(\mathbf{r}) \, ,
\label{eq:ti_gpe}
\end{equation}
while the fluctuations satisfy the linearized Bogoliubov-de Gennes equation
\begin{equation}
i\hbar\partial_t \delta \hat{\Psi}(\mathbf{r},t)
= \left[ h_{\mathrm{SO}} - \mu + h_D(\mathbf{r}) \right] \delta \hat{\Psi}(\mathbf{r},t) + h_C(\mathbf{r}) [\delta \hat{\Psi}^\dagger(\mathbf{r},t)]^T \, .
\label{eq:lin_bogo}
\end{equation}
In this equation we introduced the quantities
\begin{align}
h_D(\mathbf{r}) = {} & {}
g_{dd} [\Psi_0^\dagger(\mathbf{r})\Psi_0(\mathbf{r}) + \Psi_0(\mathbf{r}) \otimes \Psi_0^\dagger(\mathbf{r})]
+ g_{ss} \{[\Psi_0^\dagger(\mathbf{r}) \sigma_z \Psi_0(\mathbf{r})] \sigma_z + [\sigma_z\Psi_0(\mathbf{r})] \otimes [\sigma_z\Psi_0(\mathbf{r})]^\dagger \} \, ,
\label{eq:hD} \\
h_C(\mathbf{r}) = {} & {}
g_{dd} \Psi_0(\mathbf{r}) \otimes \Psi_0^T(\mathbf{r}) + g_{ss}  [\sigma_z\Psi_0(\mathbf{r})] \otimes [\sigma_z\Psi_0(\mathbf{r})]^T \, ,
\label{eq:hC}
\end{align}
and $\otimes$ is the standard Kronecker product of spinors.

\section{The stripe phase}
\label{sec:stripe_phase}
The zero-temperature phase diagram of the SO-coupled BEC described by Hamiltonian~(\ref{eq:H}) has been the subject of several works~\cite{Ho2011,Li2012PRL,Li2013}.
Its full determination requires one to calculate the ground state of the system as a function of the spin-orbit parameters $k_R$, $\Omega_R$, the couplings $g_{dd}$, $g_{ss}$,
and the average density $\bar{n} = N/V$. In the mean-field treatment that we are employing, the ground state corresponds to the solution of the GP equation~(\ref{eq:ti_gpe}),
which has the lowest energy (we recall that the energy $E_0$ of the mean-field configuration $\Psi_0$ is obtained by setting $\hat{\Psi} = \Psi_0$ in Eq.~(\ref{eq:H})
and carrying out the spatial integration).

One of the most prominent features of the phase diagram of a SO-coupled BEC is the presence of the so-called stripe phase. The wavefunction of this phase can be written
as~\cite{Li2013}
\begin{equation}
\Psi_0(\mathbf{r}) = e^{i k_c x} \sum_{\bar{m} \in \mathbb{Z}} \tilde{\Psi}_{\bar{m}} e^{2 i \bar{m} k_1 x} \, .
\label{eq:stripe_wf}
\end{equation}
This expression is a Bloch wave with quasimomentum $\hbar\mathbf{k}_c = \hbar k_c \hat{\mathbf{e}}_x$, in which the role of the reciprocal lattice vectors is played
by the quantities $\left\{ 2 \bar{m} \mathbf{k}_1 \right\}_{\bar{m} \in \mathbb{Z}}$, where $\mathbf{k}_1 = k_1 \hat{\mathbf{e}}_x$. The two-component spinor coefficients
of the Bloch expansion are denoted as $\tilde{\Psi}_{\bar{m}}$. One can easily show that for $\delta_R = 0$ and $g_{\uparrow\uparrow} = g_{\downarrow\downarrow}$,
the equalities $k_c = k_1$ and $\tilde{\Psi}_{-\bar{m}} = (\sigma_x \tilde{\Psi}_{\bar{m}-1})^*$ hold, which yields a vanishing magnetic polarization
$\langle \sigma_z \rangle = \int_V d^3r \, \Psi_0^\dagger(\mathbf{r}) \sigma_z \Psi_0(\mathbf{r})$~\cite{Ho2011,Li2012PRL,Li2013}.

In order to determine the values of the parameters entering the wavefunction~(\ref{eq:stripe_wf}) one can proceed as follows. First, one calculates the energy of the mean-field
state~(\ref{eq:stripe_wf}) as a function of the momenta $k_c$ and $k_1$ and the components of the spinors $\tilde{\Psi}_{\bar{m}}$. Then, one minimizes
the resulting expression with respect to all these quantities. In performing this procedure we have to take into account the normalization condition for $\Psi_0$, which yields
the constraint $\sum_{\bar{m} \in \mathbb{Z}} \tilde{\Psi}_{\bar{m}}^\dagger \tilde{\Psi}_{\bar{m}} = \bar{n}$. It is easy to check that the wavefunction determined in this way
is an exact solution of the GP equation~(\ref{eq:ti_gpe}). On the other hand, in the numerical calculations it is necessary to truncate the infinite sum in Eq.~(\ref{eq:stripe_wf})
to a finite number of terms. In this respect, we point out that the largest contributions to $\Psi_0$ are those with $\bar{m} = -1,0$ (retaining only these two terms reproduces the
variational ansatz employed in Ref.~\cite{Li2012PRL}), and the magnitude of the components of $\tilde{\Psi}_{\bar{m}}$ decreases with increasing $|\bar{m}|$. In the present work
we have retained $16$ terms ($-8 \leq \bar{m} \leq 7$). No significant changes have been observed when further extending these limits.

Qualitatively, the stripe phase can be regarded as a macroscopic occupation of an equal-weighted superposition of the two states lying at the minima of the single-particle dispersion.
However, because of the nonlinear terms of the GP equation~(\ref{eq:ti_gpe}), higher-order harmonic terms with wave vectors $\pm 3 k_1, \, \pm 5 k_1, \, \ldots$ have to be included
in the wavefunction~(\ref{eq:stripe_wf}). Notice also that the interaction shifts the momentum $k_1$ from the single-particle value $k_1^0$ of Eq.~(\ref{eq:k1_sp})~\cite{Li2012PRL}.

The stripe phase emerges only if $g_{ss} > 0$ in the competition between the density-density and spin-spin interaction terms in Hamiltonian~(\ref{eq:H}).
For sufficiently low values of $\Omega_R$ the spin interaction prevails, favoring an unpolarized configuration at the cost of creating modulations in the total density $n_0(\mathbf{r})
= \Psi_0^\dagger(\mathbf{r}) \Psi_0(\mathbf{r})$ along the $x$ direction, with wavelength $\pi / k_1$. The creation of such modulations entails spontaneous breaking
of the translational symmetry of Hamiltonian~(\ref{eq:H}), and it represents a clear signature of the supersolid character of the stripe phase. More specifically, the presence of a
complex order parameter ensures the existence of the superfluid current, and at the same time the presence of spatial periodicity means that the body is a crystal.

The fluctuation term of the field operator~(\ref{eq:field_decomp}) in the stripe phase can be written as~\cite{Li2013}
\begin{equation}
\delta \hat{\Psi}(\mathbf{r},t)
= \sum_{\ell,\mathbf{k} \in \mathrm{BZ}} \left[ U_{\ell,\mathbf{k}}(\mathbf{r}) \hat{b}_{\ell,\mathbf{k}} e^{- i \omega_{\ell,\mathbf{k}} t}
+ V_{\ell,\mathbf{k}}^*(\mathbf{r}) \hat{b}_{\ell,\mathbf{k}}^\dagger e^{i \omega_{\ell,\mathbf{k}} t} \right] \, ,
\label{eq:field_fluct}
\end{equation}
where $\hat{b}_{\ell,\mathbf{k}}$ ($\hat{b}_{\ell,\mathbf{k}}^\dagger$) are the annihilation (creation) operators of a quasiparticle with quasimomentum $\hbar\mathbf{k}$
and energy $\hbar\omega_{\ell,\mathbf{k}}$, and the index $\ell$ labels different bands of the excitation spectrum (see the discussion below). The two-component spinor
functions $U_{\ell,\mathbf{k}}(\mathbf{r})$ and $V_{\ell,\mathbf{k}}(\mathbf{r})$ are the Bogoliubov amplitudes obeying the ortho-normalization condition $\int_V d^3r
[U_{\ell,\mathbf{k}}^\dagger(\mathbf{r}) U_{\ell',\mathbf{k}'}(\mathbf{r}) - V_{\ell,\mathbf{k}}^\dagger(\mathbf{r}) V_{\ell',\mathbf{k}'}(\mathbf{r})] = \delta_{\ell\ell'}
\delta_{\mathbf{k}\mathbf{k}'}$. Notice that the summation in Eq.~(\ref{eq:field_fluct}) is restricted to the quasimomenta with the $x$ component in the first Brillouin zone (BZ),
i.e., $0 \leq k_x \leq 2 k_1$.

Inserting Eq.~(\ref{eq:field_fluct}) into Eq.~(\ref{eq:lin_bogo}) and equating the terms that have the same oscillatory behavior in time, one finds an eigenvalue equation
for the Bogoliubov frequencies and amplitudes:
\begin{equation}
\begin{pmatrix}
h_{\mathrm{SO}} - \mu + h_D(\mathbf{r}) & h_C(\mathbf{r}) \\
- h_C^*(\mathbf{r}) & - (h_{\mathrm{SO}} - \mu + h_D(\mathbf{r}))^*
\end{pmatrix}
\begin{pmatrix}
U_{\ell,\mathbf{k}}(\mathbf{r}) \\
V_{\ell,\mathbf{k}}(\mathbf{r})
\end{pmatrix}
=
\hbar\omega_{\ell,\mathbf{k}}
\begin{pmatrix}
U_{\ell,\mathbf{k}}(\mathbf{r}) \\
V_{\ell,\mathbf{k}}(\mathbf{r})
\end{pmatrix}
\, .
\label{eq:eig_bogo}
\end{equation}
The solutions of Eq.~(\ref{eq:eig_bogo}) can be expressed as Bloch waves of the form
\begin{align}
U_{\ell,\mathbf{k}}(\mathbf{r}) = {} & {}
e^{i \mathbf{k} \cdot \mathbf{r}} e^{i k_c x} \sum_{\bar{m} \in \mathbb{Z}} \tilde{U}_{\ell,\mathbf{k}+2\bar{m}\mathbf{k}_1} e^{2 i \bar{m} k_1 x} \, ,
\label{eq:bogo_u} \\
V_{\ell,\mathbf{k}}(\mathbf{r}) = {} & {}
e^{i \mathbf{k} \cdot \mathbf{r}} e^{- i k_c x} \sum_{\bar{m} \in \mathbb{Z}} \tilde{V}_{\ell,\mathbf{k}+2\bar{m}\mathbf{k}_1} e^{2 i \bar{m} k_1 x} \, .
\label{eq:bogo_v}
\end{align}
We then substitute solutions~(\ref{eq:bogo_u}) and~(\ref{eq:bogo_v}) into Eq.~(\ref{eq:eig_bogo}). Equating the terms of this equation, which oscillate in space with the same
wavelength, we turn it into an infinite set of coupled algebraic linear equations involving the two-component expansion coefficients $\tilde{U}_{\ell,\mathbf{k}+2\bar{m}\mathbf{k}_1}$,
and $\tilde{V}_{\ell,\mathbf{k}+2\bar{m}\mathbf{k}_1}$, as well as the Bogoliubov spectrum $\omega_{\ell,\mathbf{k}}$. In order to write down this set in a compact form,
we define two infinite-dimensional column vectors, $\mathbf{U}_{\ell,\mathbf{k}} = ( \cdots \, \tilde{U}_{\ell,\mathbf{k}+2(\bar{m}-1)\mathbf{k}_1}^T \,\,\,
\tilde{U}_{\ell,\mathbf{k}+2\bar{m}\mathbf{k}_1}^T \,\,\, \tilde{U}_{\ell,\mathbf{k}+2(\bar{m}+1)\mathbf{k}_1}^T \, \cdots)^T$ and $\mathbf{V}_{\ell,\mathbf{k}} =
( \cdots \, \tilde{V}_{\ell,\mathbf{k}+2(\bar{m}-1)\mathbf{k}_1}^T \,\,\, \tilde{V}_{\ell,\mathbf{k}+2\bar{m}\mathbf{k}_1}^T \,\,\, \tilde{V}_{\ell,\mathbf{k}+2(\bar{m}+1)\mathbf{k}_1}^T \, \cdots)^T$.
Notice that from the above ortho-normalization condition for $U_{\ell,\mathbf{k}}(\mathbf{r})$ and $V_{\ell,\mathbf{k}}(\mathbf{r})$, it follows that
$\mathbf{U}_{\ell',\mathbf{k}'}^\dagger \mathbf{U}_{\ell,\mathbf{k}} - \mathbf{V}_{\ell',\mathbf{k}'}^\dagger \mathbf{V}_{\ell,\mathbf{k}}
= V^{-1} \delta_{\ell\ell'} \delta_{\mathbf{k}\mathbf{k}'}$. Then, after performing the above procedure, we get the eigenvalue equation
\begin{equation}
\begin{pmatrix}
\mathcal{B}^{SO}(\mathbf{k}) - \mu + \mathcal{B}^{D} & \mathcal{B}^{C} \\
- \tilde{\mathcal{B}}^{C} & - (\tilde{\mathcal{B}}^{SO}(\mathbf{k}) - \mu + \tilde{\mathcal{B}}^{D})
\end{pmatrix}
\begin{pmatrix}
\mathbf{U}_{\ell,\mathbf{k}} \\
\mathbf{V}_{\ell,\mathbf{k}}
\end{pmatrix}
=
\hbar\omega_{\ell,\mathbf{k}}
\begin{pmatrix}
\mathbf{U}_{\ell,\mathbf{k}} \\
\mathbf{V}_{\ell,\mathbf{k}}
\end{pmatrix}
\, ,
\label{eq:eig_bogo_inf}
\end{equation}
where we introduced matrices $\mathcal{B}$ and $\tilde{\mathcal{B}}$, which have the entries
\begin{align}
&{} \mathcal{B}_{\bar{m}_1 \bar{m}_2}^{SO}(\mathbf{k}) =
\left[ \frac{\hbar^2}{2m} \left(k_x + k_c + 2 \bar{m}_1 k_1 - k_R \sigma_z\right)^2 + \frac{\hbar^2 k_\perp^2}{2 m}
+ \frac{\hbar\Omega_R}{2} \, \sigma_x \right] \delta_{\bar{m}_1 \bar{m}_2} \, ,
\label{eq:BSO} \\
&{} \mathcal{B}_{\bar{m}_1 \bar{m}_2}^{D} =
\sum_{\bar{m}, \bar{m}' \in \mathbb{Z}}
\left\{ g_{dd} \, (\tilde{\Psi}_{\bar{m}'}^\dagger \tilde{\Psi}_{\bar{m}} + \tilde{\Psi}_{\bar{m}} \otimes \tilde{\Psi}_{\bar{m}'}^\dagger)
+ g_{ss} \, [ (\tilde{\Psi}_{\bar{m}'}^\dagger \sigma_z \tilde{\Psi}_{\bar{m}}) \sigma_z + (\sigma_z \tilde{\Psi}_{\bar{m}}) \otimes (\sigma_z \tilde{\Psi}_{\bar{m}'})^\dagger ]
\right\} \delta_{\bar{m}_1-\bar{m}_2,\bar{m}-\bar{m}'} \, ,
\label{eq:BD} \\
&{} \mathcal{B}_{\bar{m}_1 \bar{m}_2}^{C} =
\sum_{\bar{m}, \bar{m}' \in \mathbb{Z}}
\left\{ g_{dd} \, (\tilde{\Psi}_{\bar{m}} \otimes \tilde{\Psi}_{-\bar{m}'}^T)
+ g_{ss} \, [ (\sigma_z \tilde{\Psi}_{\bar{m}}) \otimes (\sigma_z \tilde{\Psi}_{-\bar{m}'})^T ]
\right\} \delta_{\bar{m}_1-\bar{m}_2,\bar{m}-\bar{m}'} \, ,
\label{eq:BC}
\end{align}
and $\tilde{\mathcal{B}}_{\bar{m}_1 \bar{m}_2}^{SO}(\mathbf{k}) = \mathcal{B}_{-\bar{m}_1, -\bar{m}_2}^{SO}(-\mathbf{k})$, $\tilde{\mathcal{B}}_{\bar{m}_1 \bar{m}_2}^{D} = \mathcal{B}_{-\bar{m}_1, -\bar{m}_2}^{D*}$, $\tilde{\mathcal{B}}_{\bar{m}_1 \bar{m}_2}^{C} = \mathcal{B}_{-\bar{m}_1, -\bar{m}_2}^{C*}$. Notice that each entry of the
$\mathcal{B}$ and $\tilde{\mathcal{B}}$ matrices is, in turn, a $2 \times 2$ matrix acting on the two-component spinors $\tilde{U}_{\ell,\mathbf{k}+2\bar{m}\mathbf{k}_1}$
and $\tilde{V}_{\ell,\mathbf{k}+2\bar{m}\mathbf{k}_1}$. By solving Eq.~(\ref{eq:eig_bogo_inf}) one can finally determine the frequencies $\omega_{\ell,\mathbf{k}}$ of all
excited modes, as well as the corresponding Bogoliubov amplitudes~(\ref{eq:bogo_u}) and~(\ref{eq:bogo_v}). Like in the calculation of the wavefunction~(\ref{eq:stripe_wf}),
here it is also necessary to truncate the infinite set of equations~(\ref{eq:eig_bogo_inf}). In this work we have retained only the entries of the matrices $\mathcal{B}$ and
$\tilde{\mathcal{B}}$ with indices $-8 \leq \bar{m}_{1,2} \leq 7$.

\begin{figure}
\centering
\includegraphics[scale=1]{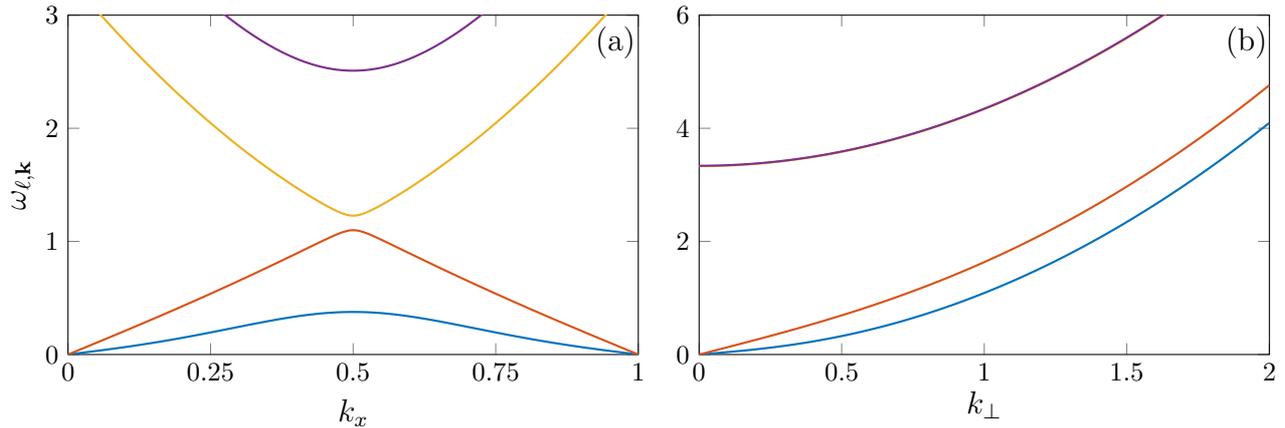}
\caption{Lowest-lying bands of the excitation spectrum of a SO-coupled BEC in the stripe phase, for excitations propagating along the $x$ axis (a) and in the transverse
$y$-$z$ plane (b). The quasimomenta $k_x$ and $k_\perp$ are in units of $2 k_1$ and $k_R$, respectively. The excitation frequencies $\omega_{\ell,\mathbf{k}}$
are in units of $E_R/\hbar$. The parameters are $\hbar\Omega_R / E_R = 2.0$, $g_{dd}\bar{n} / E_R = 0.4$, and $g_{ss}\bar{n} / E_R = 0.1$.}
\label{fig:stripe_Bogo_spectrum}
\end{figure}

The Bogoliubov spectrum of the stripe phase was originally calculated in Ref.~\cite{Li2013}. It is plotted in Fig.~\ref{fig:stripe_Bogo_spectrum} for excitations propagating along
the $x$ axis [Fig.~\ref{fig:stripe_Bogo_spectrum}(a)] and in the transverse $y$-$z$ plane [Fig.~\ref{fig:stripe_Bogo_spectrum}(b)]. The spectrum has a band structure
with two gapless branches, corresponding to two Goldstone modes originating from spontaneous breaking of the gauge and translational symmetries. At low $k$
these two branches exhibit a linear dispersion. The corresponding sound velocity is anisotropic, with minimum and maximum values for excitations
propagating along $x$ and in the $y$-$z$ plane, respectively. Furthermore, in the case of Fig.~\ref{fig:stripe_Bogo_spectrum}(a) one can see that the frequency of the two
gapless bands vanishes when $k_x$ approaches $2 k_1$, i.e., at the edge of the first Brillouin zone. Similar double gapless band structures have been found
for soft-core bosons~\cite{Saccani2012,Kunimi2012,Macri2013} and, more recently, in SO-coupled Bose gases with pure Rashba coupling~\cite{Liao2018}.

The excitation frequencies $\omega_{\ell,\mathbf{k}}$ are well defined even if $\mathbf{k}$ does not belong to the first Brillouin zone. In this case, using the properties
$\mathcal{B}_{\bar{m}_1+\bar{m},\bar{m}_2+\bar{m}}^{SO}(\mathbf{k}) = \mathcal{B}_{\bar{m}_1 \bar{m}_2}^{SO}(\mathbf{k}+2\bar{m}\mathbf{k}_1)$ and
$\mathcal{B}_{\bar{m}_1+\bar{m},\bar{m}_2+\bar{m}}^{D,C} = \mathcal{B}_{\bar{m}_1 \bar{m}_2}^{D,C}$, and similar ones for $\tilde{\mathcal{B}}$, from Eq.~(\ref{eq:eig_bogo_inf})
we obtain the periodicity of the Bogoliubov spectrum: $\omega_{\ell,\mathbf{k}+2\bar{m}\mathbf{k}_1} = \omega_{\ell,\mathbf{k}}$.

Once the Bogoliubov spectrum and amplitudes are known, one can rewrite Hamiltonian~(\ref{eq:H}) in the diagonal form,
\begin{equation}
\hat{H} = E_{\mathrm{GS}} + \sum_{\ell,\mathbf{k} \in \mathrm{BZ}} \hbar\omega_{\ell,\mathbf{k}} \hat{b}_{\ell,\mathbf{k}}^\dagger \hat{b}_{\ell,\mathbf{k}} \, ,
\label{eq:H_Bogo}
\end{equation}
where only terms up to the second order in the quasiparticle annihilation and creation operators should be retained. The constant $E_{\mathrm{GS}}$ is the ground-state energy,
which is given by the sum of the mean-field contribution $E_0$ and the so-called Lee-Huang-Yang correction due to quantum fluctuations~\cite{Lee1957}. This correction is
usually small in dilute 3D Bose gases, being proportional to $(\bar{n} a^3)^{1/2}$ for a single-component BEC with scattering length $a$. Its evaluation for our case is left for
future investigations.

We conclude the present section by briefly discussing the rest of the phase diagram of our SO-coupled Bose gas~\cite{Ho2011,Li2012PRL}. As one increases the Raman
coupling $\hbar\Omega_R$, the amplitude of the density modulations of the stripe phase grows, making their energetic cost higher and higher. Eventually, the system
undergoes a first-order transition to the so-called plane-wave phase, in which the atoms condense in a single plane-wave state with momentum $k_1$ and magnetic
polarization $\langle \sigma_z \rangle = N k_1 / k_R$. This state has the same energy as the one with opposite values of the momentum and $\langle \sigma_z \rangle$.
The critical Raman coupling at which the transition occurs takes the density-independent value $\hbar\Omega_R^{S-P} = 4 E_R \sqrt{2 g_{ss} / (g_{dd} + 2 g_{ss})}$
as $\bar{n} \to 0$~\cite{Ho2011}. With further increasing $\Omega_R$, one gets another transition, this time of the second order. This is the transition to the zero-momentum
phase with vanishing $\langle \sigma_z \rangle$, and it takes place at $\hbar\Omega_R^{P-Z} = 2(2 E_R - g_{ss} \bar{n})$.

\section{Drag force and energy dissipation in the stripe phase}
\label{sec:drag_force}
Let us now turn to the study of the motion of an impurity immersed in a SO-coupled condensate in the stripe phase. For simplicity, we consider a heavy pointlike impurity,
which has velocity $\mathbf{v}$ and is weakly coupled to the spin-$\uparrow$ and spin-$\downarrow$ components of the BEC, with equal atom-impurity $s$-wave scattering lengths $b$
and coupling strengths $g_{\mathrm{imp}} = 2 \pi \hbar^2 b / m$. The effect of the impurity can be taken into account by adding an external potential $U_{\mathrm{imp}}(\mathbf{r},t)
= g_{\mathrm{imp}} \delta(\mathbf{r}-\mathbf{v}t)$ to the single-particle Hamiltonian~(\ref{eq:sp_hamiltonian}). Within the accuracy of the Bogoliubov approach, the corresponding
contribution to the many-body Hamiltonian~(\ref{eq:H}) is given by
\begin{equation}
\hat{H}_{\mathrm{imp}}(t) = \int_V d^3 r \, \hat{\Psi}^\dagger(\mathbf{r}) U_{\mathrm{imp}}(\mathbf{r},t) \hat{\Psi}(\mathbf{r})
= g_{\mathrm{imp}} n_0(\mathbf{v} t) + \frac{g_{\mathrm{imp}}}{V} \sum_{\mathbf{q}} e^{i \mathbf{q} \cdot \mathbf{v} t} \delta\hat{\rho}_{\mathbf{q}} \, ,
\label{eq:H_imp}
\end{equation}
where $\delta\hat{\rho}_{\mathbf{q}} = \int_V d^3 r \, e^{- i \mathbf{q} \cdot \mathbf{r}} \left[ \Psi_0^\dagger(\mathbf{r}) \delta\hat{\Psi}(\mathbf{r}) + \mathrm{H.c.} \right]$
is the $\mathbf{q}$-component of the density fluctuation operator (here we switch back to the Schr\"{o}dinger representation). In some of the formulas below it will be convenient
to decompose the momentum $\mathbf{q}$ into the sum of a quasimomentum belonging to the first Brillouin zone and a reciprocal lattice vector, $\mathbf{q} = \mathbf{k}_{\mathbf{q}}
+ 2 \bar{m}_{\mathbf{q}} \mathbf{k}_1$. This can be accomplished by taking $k_{\mathbf{q},x} = 2 k_1 \left\{ q_x / 2 k_1 \right\}$, $k_{\mathbf{q},y} = q_y$, $k_{\mathbf{q},z} = q_z$,
and $\bar{m}_{\mathbf{q}} = \left[ q_x / 2 k_1 \right]$, where we have adopted the standard notation $\left\{ \ldots \right\}$ and $\left[ \ldots \right]$ for the fractional and
integer parts of the quantities, respectively. Using Eq.~(\ref{eq:field_fluct}) the density fluctuation operator becomes
\begin{equation}
\delta\hat{\rho}_{\mathbf{q}}
= \sum_\ell \left( f_{\ell,\mathbf{q}} \hat{b}_{\ell,\mathbf{k}_{\mathbf{q}}} + f_{\ell,-\mathbf{q}}^* \hat{b}_{\ell,2\mathbf{k}_1-\mathbf{k}_{\mathbf{q}}}^\dagger \right) \, ,
\label{eq:dens_fluct}
\end{equation}
where
\begin{equation}
\begin{split}
f_{\ell,\mathbf{q}} &{} = \int_V d^3 r \, e^{- i \mathbf{q} \cdot \mathbf{r}} \sum_{\mathbf{k} \in \mathrm{BZ}}
\left[ \Psi_0^\dagger(\mathbf{r}) U_{\ell,\mathbf{k}}(\mathbf{r}) + V_{\ell,\mathbf{k}}^T(\mathbf{r}) \Psi_0(\mathbf{r}) \right] \\
&{} = V \sum_{\bar{m} \in \mathbb{Z}}
\left( \tilde{\Psi}_{\bar{m}}^\dagger \tilde{U}_{\ell,\mathbf{k}_{\mathbf{q}}+2(\bar{m}+\bar{m}_{\mathbf{q}})\mathbf{k}_1}
+ \tilde{V}_{\ell,\mathbf{k}_{\mathbf{q}}+2(\bar{m}+\bar{m}_{\mathbf{q}})\mathbf{k}_1}^T \tilde{\Psi}_{-\bar{m}} \right)
\end{split}
\label{eq:dens_fluct_elem}
\end{equation}
is the matrix element of $\delta\hat{\rho}_{\mathbf{q}}$ between the ground state and the excited mode with quantum numbers $\ell$ and $\mathbf{k}_{\mathbf{q}}$.

According to the linear response theory, the time-averaged energy dissipation rate of the moving impurity is~\cite{Pitaevskii_Stringari_book}
\begin{equation}
W = - \frac{g_{\mathrm{imp}}^2}{V^2} \sum_{\mathbf{q}} (\mathbf{q} \cdot \mathbf{v}) \chi''(\mathbf{q},\omega = \mathbf{q} \cdot \mathbf{v}) \, ,
\label{eq:diss_rate}
\end{equation}
where $\chi''(\mathbf{q},\omega)$ is the imaginary part of the density dynamic response, with $\omega$ being the frequency of an external perturbation. Equation~(\ref{eq:diss_rate})
can be written as $W = \mathbf{F} \cdot \mathbf{v}$, where $\mathbf{F}$ is the time-averaged drag force that is exerted on the impurity by the condensate~\cite{Astrakharchik2004}.
In order to calculate $\mathbf{F}$ explicitly, we first recall that $\chi''$ satisfies the identity $\chi''(\mathbf{q},\omega) = \pi (S(\mathbf{q},\omega) - S(-\mathbf{q},-\omega))$,
where $S(\mathbf{q},\omega)$ is the dynamic structure factor. At zero temperature the latter is given by $S(\mathbf{q},\omega) = \hbar^{-1} \sum_{\ell} | f_{\ell,\mathbf{q}} |^2
\delta(\omega - \omega_{\ell,\mathbf{q}})$. We thus arrive at the following formula for the drag force:
\begin{equation}
\mathbf{F} =
- \frac{2 \pi g_{\mathrm{imp}}^2}{\hbar V^2} \sum_{\ell,\mathbf{q}} \mathbf{q} \, | f_{\ell,\mathbf{q}} |^2 \,
\delta\left( \omega_{\ell,\mathbf{q}} - \mathbf{q} \cdot \mathbf{v} \right) \, .
\label{eq:drag_force}
\end{equation}
An interesting consequence of the structure of Eq.~(\ref{eq:drag_force}) is the following. The contribution of the $\ell$-th branch of the Bogoliubov spectrum can be non-vanishing
only if the speed $v$ exceeds the critical value
\begin{equation}
v_{c,\ell}(\hat{\mathbf{v}}) = \min_{\mathbf{q} \cdot \hat{\mathbf{v}} > 0} \frac{\omega_{\ell,\mathbf{q}}}{\mathbf{q} \cdot \hat{\mathbf{v}}} \, ,
\label{eq:anis_critical velocity}
\end{equation}
where $\hat{\mathbf{v}}$ is a unit vector along $\mathbf{v}$ identifying the direction of the motion of the impurity. Equation~(\ref{eq:anis_critical velocity}) coincides with
a generalized version of the Landau criterion for superfluidity that takes into account the anisotropy of the system under consideration~\cite{Yu2017}. For an isotropic
superfluid it reduces to the traditional form $v_{c,\ell} = \min_{\mathbf{q}} \omega_{\ell,\mathbf{q}} / q$ independent of $\hat{\mathbf{v}}$~\cite{Pitaevskii_Stringari_book}.
If $v < v_{c,\ell}(\hat{\mathbf{v}})$ for any $\ell$, the impurity can move in the BEC along the direction $\hat{\mathbf{v}}$ without experiencing any friction.

Let us first recapitulate the results in the absence of SO coupling. In the ground-breaking work~\cite{Astrakharchik2004} Astrakharchik and Pitaevskii proved that the drag force
in a single-component 3D BEC takes a transparent expression:
\begin{equation}
\mathbf{F}_{\mathrm{SC}} = - 4 \pi \bar{n} b^2 m v^2 \left(1 - \frac{c^2}{v^2}\right)^2 \Theta(v-c) \hat{\mathbf{v}} \, ,
\label{eq:drag_force_std}
\end{equation}
where $\Theta$ denotes the Heaviside function. Hence, if $v$ exceeds the speed of sound $c$, then $\mathbf{F}_{\mathrm{SC}}$ is finite and it is antiparallel
to the velocity of the impurity. In the opposite case, $v < c$, the drag force vanishes and the motion of the impurity is dissipationless. This finding is consistent
with the Landau criterion~(\ref{eq:anis_critical velocity}), which for a standard BEC predicts a single isotropic critical velocity $v_c = c$~\cite{Pitaevskii_Stringari_book}.

All above considerations can be straightforwardly extended to miscible two-component BECs without SO coupling. In these systems the Bogoliubov spectrum is made of two branches,
a pure spin mode (the lower branch in the most typical situation $g_{ss} < g_{dd}$ that we consider in this work) and a pure density mode (upper branch). By applying the Landau
criterion~(\ref{eq:anis_critical velocity}) separately to each branch one finds two critical velocities, coinciding with the sound speeds $c_s$ and $c_d$ of the spin and density
waves, respectively. However, if the impurity has equal couplings to the two components, the spin mode does not contribute to the drag force. Hence, the final expression for
$\mathbf{F}$ remains identical to Eq.~(\ref{eq:drag_force_std}), in which $c$ is replaced with $c_d$.

In the presence of SO coupling the situation changes dramatically. First of all, since the excitation spectrum is anisotropic, from Eq.~(\ref{eq:anis_critical velocity}) it follows that
the critical speed $v_{\ell,c}$ can depend on the direction $\hat{\mathbf{v}}$ of the motion. This effect has already been addressed in several works~\cite{He2014,Liao2016,Yu2017,Kato2017},
which focused on the uniform plane-wave and zero-momentum phases. It has also been found that the drag force can be not parallel to $\hat{\mathbf{v}}$,
and the critical velocity can be different from the sound velocity even in the directions perpendicular to the condensation momentum. This is due to the emergence of
rotonlike excitations in the Bogoliubov spectrum in the plane-wave phase~\cite{Martone2012,Khamehchi2014,Ji2015}.

On the other hand, the stripe phase possesses a remarkable feature that makes it strikingly different from the others. Because of the double gapless band structure 
and of the periodicity of the excitation spectrum (see Fig.~\ref{fig:stripe_Bogo_spectrum}), the critical velocity~(\ref{eq:anis_critical velocity}) vanishes for any band $\ell$
and for any $\hat{\mathbf{v}}$ that does not lie in the $y$-$z$ plane. As a consequence, the motion of an impurity can never be dissipationless as long as it has a finite component
in the direction perpendicular to the stripes. It must be emphasized that this does not mean that the stripe phase lacks superfluidity. In fact, Eq.~(\ref{eq:anis_critical velocity})
only gives the so-called critical dragging velocity for the occurrence of a friction force on a body moving in a superfluid. In SO-coupled configurations, it does not coincide with
the critical velocity below which the system can flow with zero viscosity because of the absence of Galilean invariance~\cite{Zhu2012,Ozawa2013,Zheng2013}. A recent calculation
based on the phase-twist method has shown that the superfluid density takes a finite value in the stripe phase~\cite{Chen2018}.

Let us now calculate explicitly the drag force~(\ref{eq:drag_force}) as a function of $\mathbf{v}$. First of all, since the system possesses rotational invariance in the $y$-$z$
plane, there is no loss of generality in assuming that the projection of $\mathbf{v}$ onto this plane is directed along the $y$ axis. Moreover, since $\omega_{\ell,\mathbf{q}}$
and $|f_{\ell,\mathbf{q}}|^2$ are invariant under the inversion operation $q_i \to - q_i$ ($i = x,y,z$, and this is no longer true for the $x$ component if $\delta_R \neq 0$ or
$g_{\uparrow\uparrow} \neq g_{\downarrow\downarrow}$), one finds that taking $v_i \rightarrow - v_i$ in Eq.~(\ref{eq:drag_force}) implies $F_i \rightarrow - F_i$. Hence, we can
restrict ourselves to positive values of $v_x$ and $v_y$ and write $\mathbf{v} = v \hat{\mathbf{v}} = v \left(\cos\theta_v,\sin\theta_v,0\right)$, with $0 \leq \theta_v \leq \pi/2$.
By virtue of the above symmetry argument we immediately find that $F_z = 0$, i.e., $\mathbf{F}$ always lies in the plane spanned by $\hat{\mathbf{e}}_x$ and $\hat{\mathbf{v}}$.
Additionally, one has $F_x \leq 0$, $F_y = 0$ if $\theta_v = 0$, and $F_x = 0$, $F_y \leq 0$ if $\theta_v = \pi/2$. Hence, in these two special situations the force is antiparallel to the
velocity. In all the other cases the force is of the form $\mathbf{F} = - F \left(\cos\theta_F,\sin\theta_F,0\right)$ with $F \geq 0$ and $\theta_v - \pi/2 \leq \theta_F \leq \theta_v + \pi/2$.

We carry out the calculation of $\mathbf{F}$ by making the usual replacement $V^{-1} \sum_{\mathbf{q}} \rightarrow (2\pi)^{-3} \int d^3 q$ in Eq.~(\ref{eq:drag_force}), and
calculating the integral over the whole momentum space using cylindrical coordinates $\left(q_x,q_\perp,\varphi_q\right)$, where $\varphi_q$ is the azimuthal coordinate of
$\mathbf{q}$. The integration with respect to $\varphi_q$ is straightforward, while the one with respect to $q_x$ and $q_\perp$, together with the summation over all the bands,
has to be performed numerically. Additional details about the computation are given in Appendix~A. The results are shown in Fig.~\ref{fig:drag_force}, where we plot $F$
and $\theta_F$ as functions of the speed $v$ for various $\theta_v$ and $\Omega_R$.

\begin{figure}
\centering
\includegraphics[scale=1]{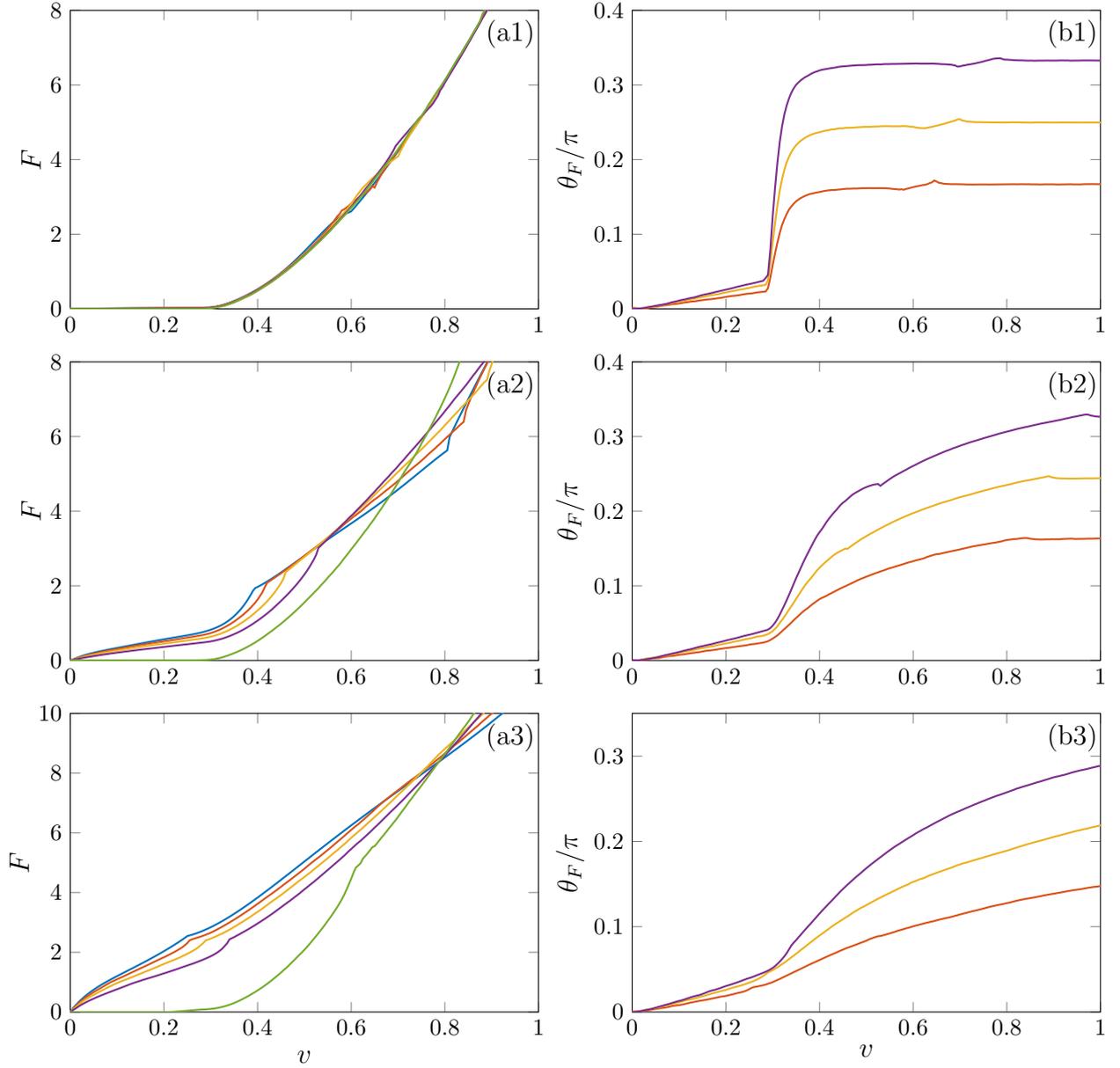}
\caption{Magnitude [(a1)--(a3)] and orientation [(b1)--(b3)] of the drag force versus the velocity of the impurity. Each couple of panels corresponds to a different value
of the Raman coupling: $\hbar\Omega_R / E_R = 0.2$ [(a1), (b1)], $1.0$ [(a2), (b2)], and $2.0$ [(a3), (b3)]. The different curves show the results for $\theta_v = 0$ (blue),
$\pi/6$ (red), $\pi/4$ (yellow), $\pi/3$ (violet), and $\pi/2$ (green). In the panels on the right column, we only display the curves for the nontrivial $\theta_v=\pi/6,\pi/4,\pi/3$ cases.
The density $\bar{n}/k_R^3 = 0.75$ and the interaction parameters $g_{dd}\bar{n}/E_R = 0.08$, $g_{ss}\bar{n}/E_R = 0.075$ correspond to those of the experiment~\cite{Li2017},
with $k_R$ increased by a factor of 2. The velocity $v$ is in units of $\hbar k_R / m$, and the force $F$ in units of $(\hbar k_R)^2 \bar{n} b^2 / m$.}
\label{fig:drag_force}
\end{figure}

As we anticipated, a finite drag force emerges at any value of the impurity speed $v$ if $\theta_v \neq \pi /2$. For low velocities, smaller than the sound speeds $c_{1,x}$,
$c_{2,x}$ of the two gapless excitation bands along $x$, the force acts mostly in the $x$ direction perpendicular to the stripes. In this regime, if $\Omega_R$ is sufficiently small,
so that the anisotropy of the sound velocity can be neglected, one finds the following expressions (see Appendix):
\begin{align}
F_x &{} \approx - \frac{16 \pi \hbar^2 k_1^2 b^2 \bar{n} |\tilde{f}_1|^2}{m^2 c_1^2} v_x \left( 1 + \frac{v_y^2}{c_1^2} \right) \, ,
\label{eq:drag_force_low_v_x} \\
F_y &{} \approx - \frac{16 \pi \hbar^2 k_1^2 b^2 \bar{n} |\tilde{f}_1|^2}{m^2 c_1^4} \, v_x^2 v_y \, ,
\label{eq:drag_force_low_v_y}
\end{align}
where $c_1$ is a typical value of the sound velocity of the lowest band, and the coefficient $|\tilde{f}_1|^2$ is coming from the matrix element $|f_{1,\mathbf{q}}|^2$.
For understanding the behavior of the drag force, it is necessary to recall that for $q_x$ close to the Brillouin point $2 k_1$ and small $q_\perp$ ($\mathbf{q}$ close to
$2 \mathbf{k}_1$) the two lowest branches of the Bogoliubov spectrum $\omega_{\ell=1,\mathbf{q}}$ and $\omega_{\ell=2,\mathbf{q}}$ acquire a strong density character.
This leads to the enhancement of the corresponding strengths, which at $\mathbf{q}$ close to $2 \mathbf{k}_1$ behave as $|f_{\ell,\mathbf{q}}|^2 \approx |\tilde{f}_\ell|^2 /
\hbar \sqrt{(q_x - 2 k_1)^2 + q_\perp^2}$ if the anisotropy of the sound velocity is negligible. Consequently, the static structure factor is also enhanced~\cite{Li2013}.
These modes with $\ell = 1,2$ at $\mathbf{q}$ close to $2 \mathbf{k}_1$ have small frequencies and can provide a finite contribution to the summation in Eq.~(\ref{eq:drag_force})
at any velocity $v$. It is also worth pointing out that typically $|\tilde{f}_1|^2 \gg |\tilde{f}_2|^2$, i.e., the enhancement of the strength of the lower gapless mode is generally
stronger than the one of the upper mode. Thus, the $\ell=1$ term in Eq.~(\ref{eq:drag_force}) is the largest at small $v$, which leads to the appearance of the factor $|\tilde{f}_1|^2$
in Eqs.~(\ref{eq:drag_force_low_v_x}) and~(\ref{eq:drag_force_low_v_y}). This is in stark contrast with the situation without SO coupling, where the lowest branch of the spectrum has
a pure spin character and is thus irrelevant for the calculation of the drag force.

The magnitude of the force increases with the Raman coupling and depends on $\theta_F$, reaching the maximum value for $\theta_F = 0$. The dependence on $\theta_v$ becomes more
pronounced at large $\Omega_R$.

When the value of $v$ grows and reaches $c_{1,x}$ and $c_{2,x}$, the low-$q$ modes of the gapless branches start to enter the summation in Eq.~(\ref{eq:drag_force}). Their contribution,
and particularly the contribution of the upper branch that has a strong density character, rapidly becomes dominant with increasing $v$. It tends to shift the orientation of the force
towards the direction opposite to the velocity. With further increasing the speed, a growing number of modes belonging to the gapped bands in the upper part of the spectrum can also
be excited by the moving impurity. In the limit of large $v$ the effects of the SO coupling become less important, and $\mathbf{F}$ is close to the value for a standard Bose gas,
which is given by Eq.~(\ref{eq:drag_force_std}) (with $c$ replaced with $c_d$).

The situation is different for $\theta_v = \pi/2$. In this case, the Landau critical velocity~(\ref{eq:anis_critical velocity}) no longer vanishes for the two gapless branches,
as it turns out to be given by the corresponding sound speeds $c_{1,\perp}$ and $c_{2,\perp}$ in the transverse $y$-$z$ plane. Hence, the dissipationless motion
of an impurity in the direction parallel to the stripes is allowed if $v < c_{1,\perp}$. This might seem surprising because for $\mathbf{q}$ in the $y$-$z$ plane
the lower and upper gapless bands are pure spin and density modes, respectively. So, one would expect no dissipation for velocities smaller than $c_{2,\perp}$, as in
the case without SO coupling. However, for all other directions of the excitation momentum $\mathbf{q}$, the lower branch has a hybrid spin and density character
due to the Raman coupling. Thus, its contribution to the drag force~(\ref{eq:drag_force}) does not vanish. A similar suppression of the dissipationless motion has been
found in the uniform plane-wave phase~\cite{Yu2017}.

The calculation of a characteristic time of the energy dissipation process requires one to know the energy of the system. If the number of impurities is $N_{\mathrm{imp}}$,
the time-averaged mean-field energy per particle reads
\begin{equation}
\varepsilon = \varepsilon_0 + \chi g_{\mathrm{imp}} \bar{n} \, .
\label{eq:av_mf_energy_imp}
\end{equation}
Here $\varepsilon_0 = E_0 / N$ is the mean-field energy per particle in the absence of impurities, and $\chi = N_{\mathrm{imp}} / N$ is the impurity concentration. We make
$\varepsilon_0$ always positive by subtracting the energy of the single-particle ground state, $\varepsilon_-(\pm\hbar\mathbf{k}_1^0) = - (\hbar\Omega_R)^2 / 16 E_R$. In order
to obtain the second term of Eq.~(\ref{eq:av_mf_energy_imp}) one has to take the time average of the mean-field contribution to the perturbation Hamiltonian~(\ref{eq:H_imp})
multiplied by $N_{\mathrm{imp}}$. Within the accuracy of our Bogoliubov treatment, the total ground-state energy of the system comprising the BEC and the impurities is given
by the sum of the mean-field energy~\eqref{eq:av_mf_energy_imp} and a correction~\cite{Astrakharchik2004}. The latter includes, besides the Lee-Huang-Yang term mentioned
at the end of Sec.~\ref{sec:stripe_phase}, an additional term proportional to $g_{\mathrm{imp}}^2$ and arising from the quantum fluctuation part of Hamiltonian~(\ref{eq:H_imp}).
However, this correction is expected to be small for the values of the parameters used in the present work. Thus, it can be safely neglected in the calculations below. 

\begin{figure}
\centering
\includegraphics[scale=1]{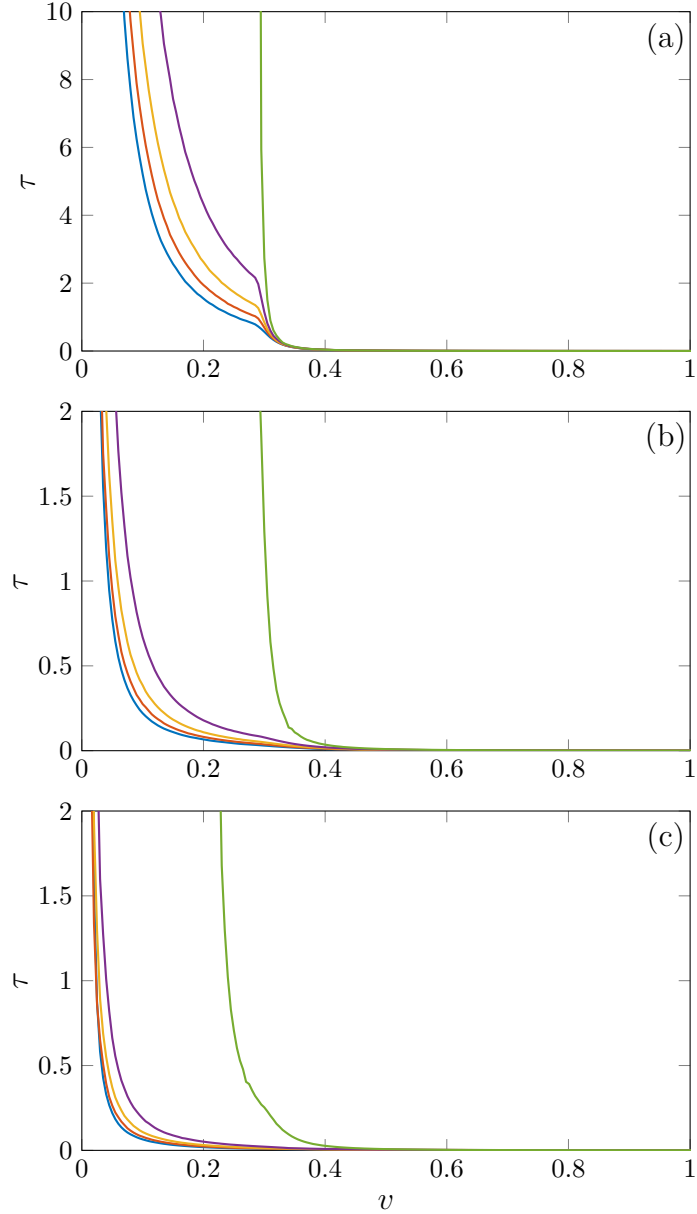}
\caption{Time scale for the energy dissipation as a function of the velocity of the impurities, for a fixed impurity concentration $\chi = 0.5$, and for $b/a_{dd} = 1.0$.
For the other parameters we use the same values as Fig.~\ref{fig:drag_force}.
Each panel corresponds to a different value of the Raman coupling: $\hbar\Omega_R / E_R = 0.2$ (a), $1.0$ (b), and $2.0$ (c).
The different curves show the results for $\theta_v = 0$ (blue), $\pi/6$ (red), $\pi/4$ (yellow), $\pi/3$ (violet), and $\pi/2$ (green).
The density $\bar{n}/k_R^3 = 0.75$ and the interaction parameters $g_{dd}\bar{n}/E_R = 0.08$, $g_{ss}\bar{n}/E_R = 0.075$ correspond to those of the experiment~\cite{Li2017},
with $k_R$ increased by a factor of 2. The velocity $v$ is in units of $\hbar k_R / m$. $\tau$ is expressed in seconds.}
\label{fig:time_scale}
\end{figure}

We define the time scale $\tau$ over which the system is superfluid as the ratio of the total energy $N \varepsilon$ to the dissipation rate $N_{\mathrm{imp}} |W|$:
\begin{equation}
\tau = \frac{\varepsilon}{\chi |W|} \, .
\label{eq:char_time}
\end{equation}
In Fig.~\ref{fig:time_scale} we plot $\tau$ as a function of $v$ for a given value of the impurity concentration $\chi$ and the ratio $b/a_{dd}$. Here $a_{dd} =
(a + a_{\uparrow\downarrow}) / 2$, and $b$, $a_{dd}$ and $\chi$ are chosen low enough to remain within the range of applicability of the mean-field approach~\cite{Astrakharchik2002}.

As expected from the behavior of the drag force discussed in Sec.~\ref{sec:drag_force}, if $v$ is much larger than the velocities of the sound modes, the energy dissipation
occurs with essentially the same features as in a BEC without SO coupling. By contrast, in the opposite regime of small $v$, the time scale becomes sensitive to the direction
of the motion $\theta_v$, increasing as it deviates from the $x$ axis and becoming infinite at $\theta_v = \pi/2$.  The time $\tau$ also exhibits a marked dependence on the Raman
coupling, reducing as the latter increases. Thus, the capability of the stripe phase to support dissipationless motion of a body becomes weaker in the presence of a well pronounced
crystalline structure. However, it is worth pointing out that, even for the largest value of $\Omega_R$ considered in this work, one has $\tau\gtrsim 0.1 \, \mathrm{s}$ for a wide
range of velocities. This is of the order of or even larger than the typical duration of experiments with ultracold atomic gases. We thus conclude that the motion of an impurity
through the stripe phase takes place, to a large extent, in the same way as in ordinary uniform superfluids.

\section{Conclusions}
\label{sec:conclusions}
We have analyzed the motion of an impurity immersed in the supersolid stripe phase of a spin-orbit-coupled Bose-Einstein condensate. After reviewing the properties of the
ground state and elementary excitations, we have calculated the drag force acting on the impurity as a function of its velocity. According to the Landau criterion for
anisotropic systems, the double gapless band structure of the Bogoliubov spectrum causes vanishing critical velocity for all directions of the motion, except for
those parallel to the stripes. At a small speed, a finite drag force arises, which is mainly associated with the excitation of the Bogoliubov modes belonging to the two lowest
branches of the spectrum and lying close to the Brillouin point. The force is not parallel to the velocity of the impurity, unless the latter is parallel or perpendicular to the
stripes. For larger speeds, low-quasimomentum modes and upper branches of the spectrum give a dominant contribution to the drag force, and our results approach the ones 
in the absence of the spin-orbit coupling. From the obtained drag force and total energy we extrapolate the time scale $\tau$ characterizing the energy dissipation process.
For parameters similar to those of Ref.~\cite{Li2017}, where the measurements were performed at low values of the Raman coupling, we find that this time scale can exceed one second,
thus ensuring that the motion of the impurity can occur with a fairly small energy dissipation for the whole duration of the experiment.

In future, it would be interesting to study the effects of the friction force on the moving striped BEC. In contrast to ordinary uniform superfluids, where it can only reduce
the velocity of the flow, in the stripe phase the friction may act in the direction of weakening or eliminating the density modulations. Future developments of the present work
may also concern the extension to finite temperatures and to other supersolid phases, such as those predicted in spin-1 SO-coupled BECs~\cite{Sun2016,Yu2016,Martone2016}
and in dilute two-dimensional dipolar Bose gases~\cite{Lu2015}.

\section*{Acknowledgments}
We acknowledge useful discussions with N. Pavloff, L. P. Pitaevskii, and S. Stringari.
The research leading to these results has received funding from the European Research Council under European Community's
Seventh Framework Programme (FP7/2007-2013 Grant Agreement No. 341197).

\section*{Appendix: Calculation of the drag force}
The calculation of the drag force is as follows. We first make the replacement $V^{-1} \sum_{\mathbf{q}} \rightarrow (2\pi)^{-3} \int d^3 q$ in Eq.~(\ref{eq:drag_force}).
Then, we switch to cylindrical coordinates by setting $\mathbf{q} = (q_x,q_\perp\cos\varphi_q,q_\perp\sin\varphi_q)$. Equation~(\ref{eq:drag_force}) then becomes
\begin{equation}
\mathbf{F} = - \frac{\hbar^3 b^2 \bar{n}}{m^2} \sum_\ell \int_{-\infty}^{+\infty} d q_x \int_{0}^{+\infty} d q_\perp \int_{0}^{2\pi} d\varphi_q
\, q_\perp \, |f_\ell(q_x,q_\perp)|^2 \delta(\omega_\ell(q_x,q_\perp) - v_x q_x - v_y q_\perp \cos\varphi_q) \, \mathbf{q} \, ,
\label{eq:drag_force_int_cyl}
\end{equation}
where we have used the fact that $\omega_{\ell,\mathbf{q}}$ and $|f_{\ell,\mathbf{q}}|^2$ do not depend on $\varphi_q$.

Let us first consider the $v_y \neq 0$ case. By using the properties of the $\delta$ function one can write
\begin{equation}
\delta(\omega_\ell(q_x,q_\perp) - v_x q_x - v_y q_\perp \cos\varphi_q)
= \frac{1}{v_y q_\perp |\sin\tilde{\varphi}_{\ell,q}|}
\sum_{\bar{m} \in \mathbb{Z}} \left[ \delta(\varphi_q - (\tilde{\varphi}_{\ell,q} + 2 \bar{m} \pi)) + \delta(\varphi_q - (- \tilde{\varphi}_{\ell,q} + 2 \bar{m} \pi)) \right] \, ,
\label{eq:drag_force_delta}
\end{equation}
where
\begin{equation}
\tilde{\varphi}_{\ell,q} = \arccos \frac{\omega_\ell(q_x,q_\perp) - v_x q_x}{v_y q_\perp} \, .
\label{eq:drag_force_phi}
\end{equation}
By plugging Eqs.~(\ref{eq:drag_force_delta}) and~(\ref{eq:drag_force_phi}) into~(\ref{eq:drag_force}) and performing the integration over $\varphi_q$, we find
\begin{align}
F_x &{} = - \frac{\hbar^3 b^2 \bar{n}}{m^2} \sum_\ell \int_{-\infty}^{+\infty} d q_x \int_{0}^{+\infty} d q_\perp
\, q_\perp \, |f_\ell(q_x,q_\perp)|^2 \frac{2\Theta(v_y q_\perp - |\omega_\ell(q_x,q_\perp) - v_x q_x|)}{\sqrt{(v_y q_\perp)^2 - [\omega_\ell(q_x,q_\perp) - v_x q_x]^2}}
\, q_x \, ,
\label{eq:drag_force_int_x} \\
F_y &{} = - \frac{\hbar^3 b^2 \bar{n}}{m^2} \sum_\ell \int_{-\infty}^{+\infty} d q_x \int_{0}^{+\infty} d q_\perp
\, q_\perp \, |f_\ell(q_x,q_\perp)|^2 \frac{2\Theta(v_y q_\perp - |\omega_\ell(q_x,q_\perp) - v_x q_x|)}{\sqrt{(v_y q_\perp)^2 - [\omega_\ell(q_x,q_\perp) - v_x q_x]^2}}
\, \frac{\omega_\ell(q_x,q_\perp) - v_x q_x}{v_y} \, ,
\label{eq:drag_force_int_y}
\end{align}
while $F_z = 0$, in agreement with the symmetry argument presented in Sec.~\ref{sec:drag_force}. The symbol $\Theta$ in Eqs.~(\ref{eq:drag_force_int_x}) and~(\ref{eq:drag_force_int_y})
denotes the Heaviside function.

All the formulas derived up to now are exact. At very low impurity speed $v$, the dominant contribution to the integrals~(\ref{eq:drag_force_int_x}) and~(\ref{eq:drag_force_int_y})
comes from the modes of the lowest-lying band of the excitation spectrum ($\ell = 1$) with $q_x$ close to $2 k_1$ and small $q_\perp$ ($\mathbf{q}$ close to $2\mathbf{k}_1$). In this
regime we can write $\omega_1(q_x,q_\perp) \approx c_1 \sqrt{(q_x-2k_1)^2+q_\perp^2}$, where $c_1$ is the sound velocity, and we neglect its anisotropy. This assumption is
reasonable when $\Omega_R$ is small and $c_{1,x} \approx c_{1,\perp}$, as shown in Ref.~\cite{Li2013}. In Ref.~\cite{Li2013} it was also proven, by means of sum-rule techniques,
that the static structure factor $S(\mathbf{q}) = N^{-1} \sum_{\ell,\mathbf{q}} |f_{\ell,\mathbf{q}}|^2$ behaves as $1/\sqrt{(q_x-2k_1)^2+q_\perp^2}$ when $\mathbf{q}$ approaches
$2\mathbf{k}_1$. Since $S(\mathbf{q})$ is dominated by the $\ell = 1$ term close to the Brillouin point, we can write $|f_1(q_x,q_\perp)|^2 \approx |\tilde{f}_1|^2 / \hbar
\sqrt{(q_x-2k_1)^2+q_\perp^2}$, where $|\tilde{f}_1|^2$ is a numerical coefficient. We now put the above expressions for $\omega_1(q_x,q_\perp)$ and $|f_1(q_x,q_\perp)|^2$ into
Eqs.~(\ref{eq:drag_force_int_x}) and~(\ref{eq:drag_force_int_y}). Then, using the polar representation $q_x = 2 k_1 + q \cos\theta_q$, $q_\perp = q \sin\theta_q$, and taking
$v_x q_x \approx 2 v_x k_1$, we arrive at the following expressions:
\begin{align}
F_x &{} \approx - \frac{\hbar^2 b^2 \bar{n}}{m^2} \int_{0}^{\pi} d\theta_q \int_0^{+\infty} d q \, q \sin\theta_q
\frac{2 |\tilde{f}_1|^2 \Theta(v_y q \sin\theta_q - |c_1 q - 2 v_x k_1|)}{\sqrt{(v_y q \sin\theta_q)^2 - (c_1 q - 2 v_x k_1)^2}} (2 k_1 + q \cos\theta_q) \, ,
\label{eq:drag_force_int_app_x} \\
F_y &{} \approx - \frac{\hbar^2 b^2 \bar{n}}{m^2} \int_{0}^{\pi} d\theta_q \int_0^{+\infty} d q \, q \sin\theta_q
\frac{2 |\tilde{f}_1|^2 \Theta(v_y q \sin\theta_q - |c_1 q - 2 v_x k_1|)}{\sqrt{(v_y q \sin\theta_q)^2 - (c_1 q - 2 v_x k_1)^2}} \frac{c_1 q - 2 v_x k_1}{v_y} \, .
\label{eq:drag_force_int_app_y}
\end{align}
Integrals~(\ref{eq:drag_force_int_app_x}) and~(\ref{eq:drag_force_int_app_y}) can be easily evaluated. It is convenient to perform first the integration with respect to $q$. The
Heaviside function reduces to unity in the range $q_-(\theta_q) < q < q_+(\theta_q)$, with $q_\pm(\theta_q) = 2 v_x k_1 / (c_1 \mp v_y \sin\theta_q)$, and vanishes otherwise.
One finally gets
\begin{align}
F_x &{} \approx - \frac{16 \pi \hbar^2 k_1^2 b^2 \bar{n} |\tilde{f}_1|^2}{m^2} \frac{v_x}{c_1^2-v_y^2} \, ,
\label{eq:drag_force_app_x} \\
F_y &{} \approx - \frac{16 \pi \hbar^2 k_1^2 b^2 \bar{n} |\tilde{f}_1|^2}{m^2} \frac{v_x^2 v_y}{(c_1^2-v_y^2)^2} \, ,
\label{eq:drag_force_app_y}
\end{align}
which become Eqs.~(\ref{eq:drag_force_low_v_x}) and~(\ref{eq:drag_force_low_v_y}) of the main text if one expands the denominators in powers of $v_y/c_1$ and retains the terms
up to cubic order in $v$.

If $v_y = 0$, the integration over $\varphi_q$ in Eq.~(\ref{eq:drag_force_int_cyl}) is trivial. It gives $F_y = F_z = 0$ and
\begin{equation}
F_x = - \frac{\hbar^3 b^2 \bar{n}}{m^2} \sum_\ell \int_{-\infty}^{+\infty} d q_x \int_{0}^{+\infty} d q_\perp
\, q_\perp \, |f_\ell(q_x,q_\perp)|^2 \delta(\omega_\ell(q_x,q_\perp) - v_x q_x) \, q_x \, .
\label{eq:drag_force_int_x_vy0}
\end{equation}
In the small-$v$ limit the integral~(\ref{eq:drag_force_int_x_vy0}) can be calculated using the same approximations as in the $v_y \neq 0$ case. The final result coincides with
Eq.~(\ref{eq:drag_force_app_x}) with $v_y = 0$.

\end{document}